\documentstyle[aaspp4]{article}

\begin{document}
 
\slugcomment{\it 24apr97 draft}
\slugcomment{\it 20jan98 draft}
\slugcomment{\it 13feb98 draft}
\slugcomment{\it 14apr98 draft}
\slugcomment{\it 23jun98 draft}
\slugcomment{\it 30jun98 draft}
\slugcomment{\it 10jul98 draft}
\slugcomment{\it 22jul98 draft}
\slugcomment{\it 05aug98 draft}
\slugcomment{\it 7 aug 98 submitted to ApJ}  
\slugcomment{\it 12 nov 98 revised}
\slugcomment{\it 11 dec 98 revised}

\title{The Supernova Remnant Cas A at Millimeter Wavelengths}

\author {Melvyn Wright\altaffilmark{1}, John Dickel\altaffilmark{2}, Barron Koralesky\altaffilmark{3}, and Lawrence Rudnick\altaffilmark{3}}
  
\altaffiltext{1}{Radio Astronomy Laboratory, University of California, Berkeley, CA 94720 E-mail: mwright@astron.berkeley.edu}
\altaffiltext{2}{Department of Astronomy, University of Illinois, 1002 West Green Street, Urbana, IL 61801 E-mail: johnd@astro.uiuc.edu}
\altaffiltext{3}{Department of Astronomy, University of Minnesota, 116 Church St SE, Minneapolis, MN 55455 E-mail: barron@astro.spa.umn.edu, larry@astro.spa.umn.edu}

\begin{abstract}

We used the BIMA array to map the supernova remnant Cas A from 28 to 87
GHz with $2''$ to $7''$ angular resolution.  Data from 75 to 87 GHz,
with 19 pointing centers were combined with single dish data to form a
completely sampled image. These new BIMA images were compared with VLA
images at 1.5 and 5 GHz to look for spectral index variations across
Cas A.  The images were spatially filtered and convolved to a common
sampled uv-range corresponding to angular scales from $7''$ to $95''$.
The images give direct evidence for a steeper spectral index in the
knots outside the bright ring.  Analysis of the 26 brightest peaks
gives statistical evidence for spectral index variations throughout the
remnant.  The high frequency spectra confirm and extend the spectral
variations seen previously at lower frequencies. The spectra are
largely consistent with different power laws, and not curved spectra.
This provides strong support for local variations in the acceleration
of relativistic particles in Cas A.

\end{abstract}
 

\section{Introduction}

Cas A has been widely studied, both as a nearby (3.4 kpc, Reed etal
1995) example of a recent (AD 1680, Ashworth 1980) supernova within our
Galaxy and also as a possible source of cosmic ray acceleration.
Braun, Gull \& Perley (1987) summarize the physical processes in Cas A
as follows. In the simplest models, a Type II SNR such as Cas A ejects
about 5 M\sun  ~ of material with velocities up to 10,000 km/s. Most of
the energy is in a small (0.3 M\sun) component at the highest
velocities.  These ejecta create a blast wave which sweeps up
interstellar material and slows to about 2300 - 3900 km/s with
densities 0.4 to 4 cm$^{-3}$. The more massive, slower ejecta are now
interacting with this shocked material creating a turbulent shell which
is seen as a bright, clumpy ring of radio emission with a radius of
about $120''$ (2pc).  A different model, where Cas A's shell is due to
the interaction of the SN explosion with a pre-existing circumstellar
shell, has been proposed by Chevalier \& Liang (1989) and refined by
Borkowski etal (1996).

Detailed multi-frequency maps of Cas A offer the possibility of
studying the particle acceleration in a diffuse plasma.  This
acceleration is thought to take place in shocks in both galactic and
extragalactic radio sources.  The synchrotron spectral index traces the
distribution of energy in relativistic particles.  In a steady state,
first-order Fermi acceleration will lead to power law spectra with a
mean spectral index of -0.5 in the presence of strong shocks (see
review by Blandford \& Eichler, 1987).  However, Cas A has a mean
spectral index of about -0.77 (Baars etal 1977), the steepest of any
known SNR. It appears that it may be flattening with time (Dent etal
1974), although this has been questioned (Hook, Duffett-Smith \&
Shakeshaft 1992).  There are several ways to produce different spectral
indices.  If the shocks are initially weak, the spectra will be steeper
than in the strong shock limit.  In addition, the shocks can be
weakened ("modified") by the backpressure of accelerated cosmic ray
ions (Jones \& Kang 1992). Radio spectra calculated for the electrons
from these cosmic ray-modified shocks show a concave shape with a
hardening toward higher energies (Reynolds \& Ellison 1992).  Green \&
Scheuer (1992) point out that, without continued acceleration, the
highest energy electrons will decay most rapidly and there should be a
break in the spectrum somewhere at high frequencies.  Turbulent
acceleration with secularly increasing magnetic fields such as
discussed by Dickel etal (1993) and Jun etal (1996) have also been
considered, and yield different spectra from first-order acceleration.
Those spectra and the efficiency of acceleration depend on local plasma
turbulence properties, (e.g., Borovsky and Eilek, 1986).

Most studies of Cas A have focussed on the bright knots, thus avoiding
most of the difficulties associated with mapping large-scale
structures, such as missing or distorted features from incomplete uv
coverage or deconvolution errors.  From a study of 304 compact knots,
Anderson \& Rudnick (1996) find variations in spectral index across Cas
A over a range from -0.65 to -0.9.  Steeper spectra are typically
associated with features suggested to be bow shocks by Braun, Gull \&
Perley (1987), and other features seen outside of the main radio ring;
flatter spectra are found in the ring and bright features within it.
The spatial scales over which significant and reliable variations are
detected range from less than $11''$ to greater than $80''$ but many
smaller scale features are present and they may have different spectral
indices as well.  These results indicate a direct link between
conditions in various parts of the remnant and the energy distributions
of the relativistic particles.

Comparison of high-resolution images at the highest frequency possible
with  maps at lower frequencies can reveal more accurately where the
high energy particles are being generated, help distinguish between the
various models, and provide a fundamental test of theories of
relativistic  particle generation in shocks.  In this paper we present
images from 28 to 87 GHz obtained with the BIMA array\footnote{The BIMA
array is operated by the Berkeley-Illinois-Maryland Association under
funding from the National Science Foundation}. We compare these new
BIMA images with  VLA\footnote{The VLA array is an instrument of the
National Radio Astronomy Observatory.  The NRAO is a facility of the
National Science Foundation, operated under cooperative agreement with
Associated Universities, Inc.} images at 1.5 and 5 GHz at similar
resolution to compute a spectral index distribution.  This study
increases the frequency baseline over which the spectra can be
determined by over an order of magnitude.

\section{Observations}

Observations were obtained with the BIMA array (Welch etal 1996) at 28,
77, and 85 GHz (see Table 1).  The primary beam of the 6.1-m diameter
antennas is Gaussian with FWHM $6.73'$, $2.49'$ and $2.25'$
respectively at 28, 77, and 85 GHz.  In order to accurately image the
full $5'$ extent of the radio source we observed on a hexagonal grid of
7 pointing centers with $2.5'$ spacing at 28 GHz, and 19 pointing
centers with $1'$ spacing at 77 and 85 GHz.  BL Lac was observed as a
phase calibrator at 25 to 30  min. intervals, after 2 complete cycles
of 19 pointings with 35s observations at each pointing at 77 and 85
GHz, and 6 complete cycles of 7 pointings at 28 GHz.  The antenna gains
were determined from observations of planets at short antenna
separations. All data were obtained in an 800 MHz bandwidth using a
digital correlator with 256 frequency channels.  The 28 GHz
observations were single sideband with a system temperature between 40
and 60 K.  The 77 and 85 GHz observations were obtained in both
sidebands of the local oscillator (LO) with single sideband system
temperature between 190 and 450 K. All system temperatures are scaled
to outside the atmosphere.

Single dish data at 87 GHz were obtained using the NRAO 12m telescope
using on-the-fly mapping in 1996 September. The single dish beam
pattern was obtained by mapping Saturn using the same technique, with a
small correction for beam broadening due to Saturn's $19.43 \times
17.36''$ size.  Saturn moved less than $3''$ during the mapping.  The
corrected beam size was determined to be $76.6 \times 72.4''$.  Maps
were made by RA scans at $25''$ intervals and after all the data were
combined into a single map, we subtracted linear baselines for each RA
scan.  The Saturn observations gave an antenna gain 31.7 Jy/K and an
integrated flux density of 96.7 Jy for Cas A for an adopted planetary
brightness 149 K.


Maps from the BIMA data were made with the MIRIAD software package
(Sault, Teuben \& Wright 1995; Wright \& Sault 1993).  The
interferometer data were combined using a maximum entropy mosaicing
algorithm (MEM) using the single dish image as a default image
(Cornwell \& Fomalont 1989; Sault, Staveley-Smith, \& Brouw 1996).  The
single dish data were deconvolved using a Gaussian beam with FWHM $77
\times 72''$. The single dish image at 87 GHz was scaled by the mean
spectral index and used in a joint deconvolution at 75 to 87 GHz.  We
first made images in each available sideband of the LO, i.e. five
images with mean frequencies of 28.5, 75.2, 78.7, 83.1, and 86.7 GHz.
Each image used multifrequency synthesis (MFS) with eight 100 MHz bands
across the 800 MHz wide sideband to reduce the effects of bandwidth
smearing over the large field of view, and combined the multiple
pointing centers into a mosaiced image which has been corrected for the
primary beam. The 75.2 and 78.7 GHz images showed structures which
differed at the $20\%$ level even after careful editing of the data.
The deconvolved images were quite sensitive to the uv-coverage when
different subsets of the data were included in the imaging. The
uv-coverage is improved by combining both sidebands of the LO at 77,
and at 85 GHz using MFS synthesis.  Although the 28.5 and 85 GHz images
were very similar, there were significant differences between these and
the 77 GHz image.  The 77 GHz image was still quite sensitive to
different subsets of the data, even with both LO sidebands combined.
This image was made using a single configuration of the BIMA array with
10 antennas (Table 1).  We are obliged to acknowledge that an MFS,
mosaiced image with 45 baselines, 16 frequency bands and 19 pointing
centers may be inadequate to look for spectral index variations in a
complex source structure like Cas A.  We therefore combined the 77 and
85 GHz data into a single MFS, mosaiced image with mean frequency 83.1
GHz, 117 baselines, 16 frequency bands and 33 pointing centers.

The 1.5 GHz images were obtained using the VLA between 1994 March 25
and 1995 March 25, with an effective (integration time weighted) epoch
1994.75 at a mean frequency 1.51 GHz.  The 5 GHz images were obtained
using the VLA at frequencies 4.4149, 4.9851, 5.0850 and 4.6351 GHz
between 1994 June 30 and 1995 March 25, with effective epoch 1995.0,
and mean frequency 4.8 GHz.  The primary beam of the 25m diameter VLA
antennas is approximately Gaussian with a FWHM $29.3'$ at 1.5 GHz, and
$9.2'$ at 4.8 GHz. These data are described in more detail by Koralesky
\& Rudnick (1998).  Maps from the VLA data were made using MEM, with
flat defaults. The use of single dish images or scaled 1.5 GHz images
as defaults, did not significantly affect the flux densities on angular
scales studied here.


\section {Results}

Figure 1 shows the maximum entropy image with a mean frequency 83.1
GHz.  The convolving half-power beamwidth was $6.5'' \times 6.2''$ .
This image includes the single dish data, and samples spatial
frequencies from zero to 72590 wavelengths.  The 83 GHz image is very
similar to the 1.5, 5 and 28.5 GHz images with the same angular
resolution.  Figure 2  shows the maximum entropy images at 1.5, 5, 28,
and 83 GHz all convolved to a common resolution of $7''$. The same
structures are apparent at all 4 frequencies.

  The spectral index of the integrated flux density is not well
determined by these data.  The total flux density at each frequency is
not measured by an interferometer, and neither the integral over the
maps nor the measured visibility flux density on the shortest baselines
(which differ for each frequency) provide good estimates of the total.
The flux density is determined mainly by the single dish observations,
which we used as a default image in the MEM deconvolution at 83 GHz,
and by the shortest interferometer spacings.  Although we could adjust
the total flux density by adding in a low resolution image from single
dish data at each frequency or by scaling a more complete image at a
lower frequency, these data are much better utilized to look for
variations with frequency in the brightness distribution across the
source.  For this purpose we do not need the absolute flux density
scale.  Indeed, the analysis is simplified if we remove the overall
spectral index (SI), and look for variations in the structure on scaled
images at each frequency.  Figure 3 shows the difference between pairs
of images after scaling to the mean spectral index (here called $\alpha_
0$) over the image.  The scale factors were determined for each pair
using the MIRIAD task IMDIFF which finds optimum parameters (in a
maximum likelihood sense) for making one image approximate another
image. The parameters are an amplitude scale factor, dc offset, shifts
in x and y direction and an expansion. For all pairs of images the
shifts were less than $0.2''$ and the dc offset and expansion less than
$0.1\%$.  The amplitude scale factor is determined globally and does
not depend on any one feature, although similar values are obtained if
we use the strongest peak to scale the images.

Significant large angular scale differences are apparent in these
images.  These may be a result of instrumental effects such as
different uv-sampling, primary beam and pointing errors, calibration
and image processing errors.  Using the difference between pairs of
images we can identify large angular scale errors which must be present
in each of the images at 1.5, 5, 28, and 83 GHz.  To avoid confusion
from these errors, we performed spectral measurements only on small
angular scales.

In order to evaluate the spectral index distribution accurately, we
must match the data as closely as possible, not only in angular
resolution but also in the spatial sampling of the images. One
possibility is to compare completely sampled images at each frequency.
For this we need complementary single dish data at each frequency which
is not currently available.  On the other hand, the SI of the knots and
filaments may differ from that of the smoother background. In this case
we probably don't want to compute the SI from completely sampled
images, but rather obtain the SI separately for filaments and
background.  Accordingly, to study the small scale structures, we
spatially filtered the data to a common range of spatial frequencies.
Since the large scale errors at 83 GHz correspond to spatial
frequencies derived from the single dish data, we limited the range of
spatial frequencies to those observed directly with interferometers.
The images were convolved with a function having a Gaussian taper at
both high and low spatial frequencies since we wish to avoid images
which are sensitive to the detailed uv-sampling at the edges.

Figure 4 shows maximum entropy images at 1.5, 5, 28, and 83 GHz each
spatially filtered to the same common uv-range defined by a circular
annulus from 1870 to 26500 wavelengths. The effective convolving beam
includes structures from $7''$ to $95''$.  These images were scaled to
the mean spectral index over the image using IMDIFF, as above.  The
mean spectral index, determined from the amplitude scale factors for
each pair of images, corresponds to a spectral index, $\alpha_0$ =
-0.76 to -0.79, where $S \propto \nu^{\alpha}$.  We also made images
with a convolving beam from $7''$ to $60''$ for comparison.  The mean
spectral index $\alpha_0$, for these images, corresponds to a spectral
index, $\alpha_0$ = -0.77 to -0.81. Although we might attribute this
steeper spectral index to the smaller scale structure, the spectra
might also be steepened by the poorer atmospheric phase coherence on
the long baselines at high frequency.

Figure 5 shows the difference between pairs of spatially filtered
images, after scaling to the mean spectral index over the image.  The
contour interval is 9 mJy/beam at 83 GHz, and scales as $\nu^{-0.77}$.
There are some common features on these images which we believe
correspond to real spectral index variations across the source. Other
features, which are not reproduced on each of the difference images may
not be real. To determine the significance of the features on the
difference images we need to consider the sources of errors on the
images.

\section {Imaging Errors}

Image errors are usually described by the thermal noise level and by
the dynamic range (the ratio of the peak brightness to the off-source
RMS).  More germane to the problem of looking for spectral index
differences across Cas A, is the image fidelity, i.e. how well does the
image represent the real source brightness distribution.  In the
process of {\it reducing} the data we re-calibrated and imaged Cas A
many times.  The RMS difference between these images was about 7
mJy/beam at 83 GHz ($1.5 \times$ the thermal noise) and 5 mJy/beam at
28 GHz ($10 \times$ the thermal noise).  The off-source RMS on the
spatially filtered images at 1.5 and 5 GHz is 11 and 9 mJy/beam,
respectively.  The off-source RMS is not well determined on the
mosaiced images at 28 and 83 GHz.  We will use these empirical
estimates of the RMS in evaluating the results.


A number of systematic errors may affect the derived spectral index
distributions: 
i) incomplete images from poor uv-coverage, 
ii) calibration and flux density scale errors,
iii) alignment and temporal changes in the images,
iv) primary beam and pointing errors,
v) deconvolution errors,
vi) polarization errors.

i)  The convolved images in Figure 2 at 1.5, 5, 28, and 83 GHz are well
sampled for structures on the scale of the individual clumps. These
images are not well sampled on angular scales greater than about $3'$
corresponding to the shortest spacing, $ 475 \lambda$ at 5 GHz, and
$600\lambda$ at 28 GHz. The 83 GHz image depends on the single dish
data for large angular scales.  The spatially filtered images (Figure
4) at 1.5, 5, 28, and 83 GHz are well sampled for structures from $7''$
to $95''$. Although the sampling is not identical, the maps should be
complete within these angular scales.

ii) The flux density scale at 28.5 and 83 GHz is tied to observations
of planets.  Flux density measurements at different observatories are
typically consistent at the 10\% level, with a worst error of 20\%.  At
85 GHz, the  two sidebands, which are separated in frequency by 3 GHz,
were independently calibrated; these images were consistent to within
10\%.  Since we determined the amplitude scale factors from the
difference images, we have sidestepped the dependence on the flux
density scale, but take encouragement from the agreement of the
spectral indices derived from the amplitude scale factors, and the
overall spectral index of Cas A (Baars etal 1977).  Calibration errors
can result in image errors and SI artifacts. The calibrator, BL Lac is
sufficiently close to Cas A, that baseline errors are not important.
Calibration errors are thought to be better than 10\% in amplitude and
$5 \deg$ in phase; both contribute around 10\% errors in the measured
visibilities.  The errors in the image are reduced by the square root
of the number of independent measurements.  Assuming a  1-2 hour time
scale for the calibration errors with two antenna configurations, gives
around 1\% errors in the images.

iii)  The alignment of the images was constrained by interpolating the
1.5, 5 and 83 GHz image onto a template using the 28 GHz image.  The
phase centers for the BIMA and VLA observations differed by $0.8^{s}$
in RA and $1.7''$ in DEC, but the data were imaged onto a common
tangent point. The observed differences in the structure are not
consistent with a simple rotation, expansion or shift of the images.

The expected proper motions and changes in source structure are also
small compared with our resolution.  The expansion rate for the bright
ring is 0.06-0.15 \% per year (Koralesky \& Rudnick 1998), leading to
maximum displacements of $0.5''$ over the 3-year span of our
observations.  The position uncertainties are estimated to be about
1/10 of the angular resolution.  The corresponding errors in the
spectral index between 83 and 5 GHz are $\pm$ 0.05 and $\pm$ 0.35 at
the peaks and 10\% contours of the clumps respectively.  In the
analysis we will discuss only the bright peaks and clearly the spectral
index on the steep edges of the clumps is to be treated with caution.

iv) The primary beam response at 28 GHz, was determined from
observations of the radio source 3C454.3, with a flux density of
approximately 8.7 Jy. The source was observed with a grid pattern of
pointing offsets over $300''$ x $300''$, with $75''$ spacing, and over
$360''$ x $360''$ with $90''$ spacing.  The best fit Gaussians for the
beam pattern were $386''$ x $380''$, and $382''$ x $379''$,
respectively, with 1\% to 2\% residuals.  We measured the primary beam
response at 90.4 GHz using a transmitter which is installed at 7 degees
elevation at a distance of 5 km. The antennas were focussed on the
transmitter and scanned across the transmitter using the
interferometer.  The measured voltage primary beam pattern was
interpolated out to $800''$ on a two dimensional grid with $48''$
spacing by using a second antenna as reference.  Within the half-power
points the primary beam response is well fitted by a Gaussian with FWHM
$127''$ at 90.4 GHz, in good agreement with the Gaussian primary beam
model used for mosaicing.  Although the primary beams for both the BIMA
array and the VLA are well determined, the primary beam models do not
perfectly represent the actual primary beam patterns.

Figure 6 shows profiles of the measured voltage primary beam pattern of
a BIMA dish using holography on a transmitter at 90.4 GHz (solid line).
The dashed line shows the Gaussian model used in the image
reconstruction, and the dot-dashed line shows the differences between
the measurement and the model. Differences as large as $20\%$ of the
model occur within the 5\% cutoff used in the image reconstruction.
Beyond the 5\% cutoff point, the real primary beam has sidelobes up to
5\% in voltage (0.25\% in power) which extend over the entire measured
pattern. This sidelobe response is not included in the primary beam
model used for mosaicing.  The sidelobe response is unrepresented, and
also contributes to amplitude errors in the visibility data.  In an
attempt to estimate the errors which might result from the difference
between the primary beam model and the real primary beam pattern we
made images using different primary beam models.  These images differed
at the 1\% level.

Pointing errors also affect the amplitude of the visibility data.  The
RMS pointing errors are about $5''$ in each axis. Although this is only
1/28 of the FWHM at 83 GHz, it gives 10\% amplitude errors in the
visibility data at the half-power points, and  20\% errors at the 5\%
point of a Gaussian primary beam response.  The errors at the 5\%
points are, of course, weighted down in the mosaiced image.  The effect
on the image fidelity depends on the source structure and on the time
scale of the pointing errors. The source structure is extended, so that
there is bright emission over most of the primary beam. The time scale
of the pointing errors is around 1 hr.  The errors in the image are
reduced by the square root of the number of independent measurements.
Assuming that all antennas have uncorrelated pointing errors results in
1\% errors on the images.

v)  Deconvolution errors can arise in several ways. A well known $\it
feature$ of maximum entropy deconvolution is the enhanced resolution of
high signal-to-noise ratio unresolved peaks in the image compared with
lower brightness peaks. Note that we have compared images which have
been convolved with a common Gaussian beam.  We also made mosaiced
images using subimages of different sizes. These mosaiced images
differed by about 0.1\%. The halo of emission around the shell in
Figure 1 is an imaging artifact arising from the inadequate
representation of the single dish beam by a Gaussian function. A better
estimate of the single dish beam shape might be obtained by
deconvolving the Saturn map using a model for Saturn, and allowing for
the rotation of the beam on the sky. However, since we have spatially
filtered out the large scale structure in deriving the Cas A spectra,
the point is moot.

vi) The BIMA images were made with linearly polarized feeds.  The
resulting images are of the total intensity, together with an error
term due to linearly polarized emission which is rotated by the
parallactic angle at each sampled visibility.  Cas A has a fractional
polarization of 5\% with an azimuthal E field when mapped at millimeter
wavelengths at $1'$ angular resolution (Kenney \& Dent 1985). We
investigated the effect on the images using a model of Cas A with an
azimuthal polarization pattern, with simulated mosaiced observations.
The resulting images using linearly polarized feeds, which are
sensitive to linearly polarized structure, differ from the true total
intensity images by 0.6\% RMS, with a maximum error of 2\%.  In
addition, polarization leakage will contaminate the total intensity
images at all frequencies at a typical level of 0.2\%.

In summary, we expect that the images have errors at the 1\% to 2\%
level, in agreement with the empirical estimates from the ensemble of
images made using in the course of the data reduction.

\section{Analysis of Spectral Index Variations}
\subsection{Direct evidence}

The spatially filtered images provide direct evidence for spectral
index differences for a few of the bright features.  For example the
clumps outside of the bright ring have a steeper spectral index than
the mean spectral index $\alpha_0$, on each of the difference maps in
figure 5.

We first look for spectral index variations on the difference maps
because the noise is better defined than on spectral index maps
involving a {\it ratio}.  The hexagonal mosaic pattern used for the
images at 28 and 83 GHz give a uniform sensitivity within a few percent
over the entire image. The RMS noise level on the difference images is
around 10 mJy/beam measured at 83 GHz, about 1 contour in Figure 5.

The flux density on the difference images is $\delta S$ = $S -
{(\nu/\nu')}^{\alpha_0} S'$, where S and $S'$ are the flux densities at
frequencies $\nu$ and $\nu'$, and $\alpha_0$ is the mean spectral index
determined from the amplitude scale factor.  For example, between 83.1
and 4.8 GHz, ${\delta{\alpha}}$ = $3.0 \times \delta S/S'$ for small
values of $\delta S/S'$.  i.e. all peaks $>$ 1 Jy at 4.8 GHz with
$\delta S <$ 10 mJy/beam on the difference maps have ${\delta{\alpha}}
< 0.03$.  Thus the brightest feature on the difference maps, with
$\delta S$ = -30 mJy/beam and $S'$ = 0.8 Jy gives ${\delta{\alpha}}$ =
-0.11.


In order to make reliable determinations of spectra in Cas A, we
selected 26 distinct features that are readily identifiable on each of
the four filtered images (Figure 7).  The flux densities for the 26
components were determined by convolving the spatially filtered maps at
each frequency by a $7''$ Gaussian beam centered on the 5 GHz peak.  At
a resolution of $7''$, we are averaging over much of the fine-scale
structure studied, e.g., by Anderson \& Rudnick (1996, AR96).  However,
they found that the largest variations in spectra occurred on large
angular scales, so we assume that we are sampling the major spectral
differences within the remnant.

The spectra for the 26 components are plotted in Figure 8, scaled by a
spectral index -0.77 (Baars etal 1977), and normalized to 83 GHz.  Fits
to the 4-point spectral index range from -0.95 to -0.75.  The spectral
indices are listed in Table 2.  The table lists non-linear least
squares fits to the flux densities at 1.5, 5, 28, and 83 GHz.  The
average and RMS for the spectral index of the 26 components is $-0.82
\pm 0.05$ The compact components are steeper on average than the
integrated flux density.  The steepest 7 spectra include the outer 5
compact components A, B, C, E, and R.

In order to test the robustness of these flux densities, we also
measured them using a second, slightly different procedure.  The 26
peak positions for each component were determined from the 5 GHz map.
A box of $10\arcsec$ centered on each peak position was then searched
in each of the other frequency maps and the peak flux density
determined at each frequency.  Both procedures allow for misalignments
due to small structural changes or motions between the maps, but differ
in the degree of large-scale filtering.  Comparison of the flux
densities for the 26 peaks between the two methods showed that the RMS
deviation in derived spectral index was 0.015 between 1.5 and 5 GHz and
0.053 between 5 and 83 GHz.

Using the errors at each frequency as discussed in section 4, we find
significant variations in spectral index among the 26 peaks.  The range
of spectral indices seen between 1.5 GHz and 83 GHz (-0.7 to -0.9) is
similar to that seen between 1.5 GHz and 5 GHz, both here and as
previously reported by AR96.  The spectral indices between 5 and 28 GHz
show a few points, typically with large errors, beween -0.9  and
-1.15;  we believe these are spurious, reflecting the poorer coverage
at 28 GHz, as discussed further below.

Since the centimeter and millimeter wavelength measurements were taken
approximately two years apart, it is possible that changes in component
flux densities over that period could distort the observed spectra.  In
order to estimate this, we determined the yearly fractional changes in
the flux densities of our components using data from 1987 and 1994.  We
then used this to calculate the apparent changes in spectral index
between 5 and 83 GHz that would be caused by such variations, if they
continued at the same rate.  The RMS change in spectral index was
0.013, much smaller than the component-to-component variations reported
here, and thus not affecting any of our conclusions.

\subsection{Color-Color plots}
Another way to display the relationship between low and high frequency
spectral indices for a collection of points is to plot them together in
a scatter plot.  Such a "color-color" diagram is shown in Figure 9.
Diagrams such as this have been used to study spectral shapes for
extragalactic synchrotron sources (e.g., Katz-Stone, Rudnick \&
Anderson 1993; Katz-Stone \& Rudnick 1997)


There are a variety of advantages to using the color-color diagram.
Spectra for a variety of locations in a source can be compared with one
another in a single diagram.  Small variations in spectral shape are
easily detectable in color-color space.  Flux density calibration
errors at one frequency show up simply as a shift of all points in
color-color space.  Finally, and perhaps most important, if spectral
variations are found within a source, one can tell whether these are
all consistent with the same spectral shape.

The color-color diagram for the 1.5, 5, and 83 GHz data is shown in
Figure 9.  The errors on each point are the larger of (a) the formal
1-sigma errors due to the noise in the input maps  or (b) one-half the
difference between the spectral indices obtained by the two different
peak flux density determination methods described above.

The low and high frequency spectral indices are clearly correlated with
each other for the sample of 26 peaks.  This confirms that the spectral
variations seen between 1.5 and 5 GHz are also present up to 83 GHz,
for the sample as a whole. This rules out mechanisms that produce
curvature at the low end of the spectrum as explanations for the
spectral variations discussed by AR96.  We note that this is an
ensemble conclusion.  Individual points do deviate from a simple power
law by as much as 0.1 in spectral index which could represent slight
spectral curvature for some compact components, or other errors such as
changes in component flux densities between the centimeter and
millimeter observations.  In addition, the data are clustered closely
around the power-law line.  The deviations from exact power-laws of
different slopes are an order of magnitude smaller than would be
expected even for models with minimal spectral curvature, such as those
with a continuous injection of fresh electrons (e.g. Pacholcyzk, 1970).
A continuous injection model with a low-frequency index of -0.75 is 
shown as the dotted line in Figure 9.  The data
are therefore not consistent with spectral variations in Cas~ A being
due to a single electron population being sampled in different magnetic
fields or having experienced different rates of high frequency losses.
The most reasonable interpretation is that the data reflect different
power-law slopes at different locations in the source;  the chi-square
for this model is 1.6 per degree of freedom (or 1.3, eliminating the
worst point).  The power laws extend at least from 1.5 GHz to 83 GHz;
curvature beyond these limits is not probed.

The color-color diagram for the 1.5, 5, and 28 GHz data is not
reliable, and is not shown here.  There is a tendency for more points
to fall below the power-law line than above, which could result either
from losses steepening the high frequency spectra or from a bias
towards low 28 GHz flux densities.  Although many points still fall
near the power-law line, there is a clear population far below the
line. We do not believe these represent a reliable measurement of
spectral shapes. If they did, then the corresponding points in the
(1.5, 5, 83) GHz color-color diagram would have to fall even further
below the power-law line, which they don't.  We therefore attribute
these deviations in the (1.5, 5, 28) GHz diagram to problems in the
reconstruction of the 28 GHz map.

Most of the steepest spectrum components show a somewhat concave
spectrum (see figure 8), which can also be seen in the color-color
diagram as the group of points above the line.  Such concave spectra
are expected from cosmic-ray modified shocks (Reynolds \& Ellison
1992).  At present, we do not have a sufficient understanding of our
errors to conclude that these deviations from exact power-law shapes
are real.  However, we note that if the millimeter wavelength flux
density time variations are in the same sense, but more rapid than
those at centimeter wavelengths, then such apparent curvature would be
caused by our measurements being separated in time.  Multi-wavelength
observations at more closely spaced epochs are needed to isolate subtle
spectral vs. temporal effects.

\section{Discussion}

In the previous section, we have established that the spectral
variations seen at centimeter wavelengths in Cas A also extend at least
to 3mm.  In addition, the spectra show very little curvature over the
range 1.5 GHz to 83 GHz, and are consistent with power laws with
indices from -0.7 to -0.9.

Although the spectral variations in Cas A have been known for some time
(Anderson etal 1991), until now it has not been possible to distinguish
among the variety of possible physical causes. First, we can now rule
out differential absorption by thermal ionized gas as the cause of
these  centimeter to millimeter wavelength spectral variations.  Such
absorption probably does affect the spectra at longer wavelengths
(Kassim etal 1995).  However, since the optical depth varies as
$\nu^{-2.1}$, variations in spectral index over the range 0.2 between
1.5 and 5 GHz would translate only to a range 0.01 between 5 and 83
GHz, much smaller than is observed.

Another possibility is that the same very low frequency spectral index
is found at all locations in Cas A (except for absorption), and that
the spectral slopes at centimeter wavelengths are the result of high
frequency breaks or other curvature in the spectra.  One way to achieve
this is through synchrotron losses, which are responsible for most of
the spectral variations seen in extragalactic sources.  The synchrotron
lifetime at frequency $\nu$ GHz in a field B Gauss $\sim$ 1.6
B$^{-1.5}$ $\nu^{-0.5}$ years.  The estimated equipartition magnetic
fields range from 1 to 5 mG in the bright knots giving a synchrotron
lifetime 5000 to 500 years, at 83 GHz.  Although the synchrotron
lifetime at 83 GHz could lead to steepening of the spectrum at high
frequency, the observed spectra do not show a spectral break.  The
synchrotron lifetimes at centimeter wavelengths are much longer than
the age of the remnant, and cannot be the cause of steeper spectra in
some locations.  It is possible to construct scenarios in which the
magnetic fields are higher than their minimum values, especially in the
earlier phases of the SNR evolution, and caused significant losses to
the electrons now radiating at centimeter wavelengths.  Such scenarios
are ruled out by the current observations, because they would lead to
spectral shapes of much greater curvature than observed.  In other
words, for any reasonably shaped synchrotron spectrum with losses, the
observed variations in the centimeter wavelength spectral indices
should lead to much larger variations between centimeter and millimeter
wavelengths than are observed.

We can also ask whether the observed spectra could be due to observing
the electrons at different locations in different stages of the
acceleration process, before they reach their asymptotic power laws.
In such a picture, the electron energy distribution should cut off at
high energies, where the particles have not had sufficient time to
accelerate, or synchrotron or adiabatic losses occur on shorter
timescales than the reacceleration (e.g., Blandford \& Ostriker 1978;
Volk \& Biermann 1988).  Detailed models with small diffusion
coefficients (e.g. Kang and Jones 1991), can produce power laws over
our limited frequency range of 30. But models with larger or
energy-dependent diffusion coefficients lead to spectra which are too
curved to be compatible with our observations, when the time scales for
electron acceleration are comparable to the age of the accelerating
shocks.

The remaining possibility is that the spectra at different positions
represent the different power laws (at least over the 1.5 - 83 GHz
range) resulting from slightly different local particle acceleration
conditions.  Such a scenario was discussed by Anderson \& Rudnick
(1996), who concluded that variations in external density and
temperature could easily lead to variations in spectra of the observed
magnitude, in a first-order Fermi accleration process where the
spectral slope depends on the shock Mach number.  However, they were
unable to find a direct relation between dynamical properties of the
compact features and their spectra. In fact, the spectral variations
seen at centimeter wavelengths are not random from feature to feature,
but are coherent over much larger scales, especially steepening outside
of the bright ring.  As argued by Anderson etal (1994), the particle
acceleration may be regulated and vary over large scales (e.g., up to 1
pc), with compact features such as discussed in this paper representing
regions of magnetic field amplification which then simply illuminate
the background electron population.


We therefore conclude that the spectral variations observed at
centimeter and millimeter wavelengths result from different local
particle acceleration conditions in this complex and rapidly evolving
young remnant. The relevant electron energies are $\approx GeV$. It is
not yet clear what the limiting energy for these processes may be.
Allen etal (1997) present evidence for a high energy X-ray tail in the
spectrum of Cas A, which may represent synchrotron electrons of
energies up to 4 x $10^7$ MeV.


The mean spectral index on the spatially filtered maps presented here
is -0.77, the same spectral index as the integrated flux density on the
unfiltered maps.  Since the spatially filtered maps contain only $18
\pm 1 $ \% of the integrated flux density on the unfiltered maps, we
conclude that either acceleration occurred in very large scale features
that are currently not visible, or particles accelerated in structures
between a size range of $95''$ to $7''$ must diffuse throughout the
entire remnant.

Since the spectral index is steeper than -0.5, under the shock
acceleration model, the entire shell is decelerated - not just the
features outside the bright ring, which are even more so. 
As argued by Anderson \& Rudnick (1996), the clumps are brighter
because of increased emission due to field amplification. Particle
acceleration could also increase the emission, but compact features
would then have different spectral indices than their
local diffuse emission, which is not observed.

The steepest 7 spectra include the 5 compact components A,B,C, E, and R
located outside the bright ring.  The steepest spectrum fitted, $\alpha
= -0.95 \pm 0.05$, was for feature C, which was found by Braun et al.
(1987) to have the characteristic structure of a bow shock.  The
particle acceleration conditions are therefore different at the bright
ring, and exterior to it, and consistent with the suggestion that these
clumps may have been decelerated so that their low Mach number allows a
steeper spectrum for the shock acceleration (Anderson \& Rudnick
1966).

We performed some simple calculations to estimate how much
homogenization of the electron populations might have occurred through
spatial diffusion. We can think of diffusion as a stochastic process
with a coherence time dt years.  We estimate the diffusion using the
Alfven velocity, $v_a=B/\sqrt{4\pi\rho}$.  Using an equipartition
magnetic field, B = 3 mG, and thermal density 2 cm$^{-3}$, then v$_a$
$\sim$ 4000 km/s (about the same as the current, decelerated knot
velocity).  Suppose that the electrons stream along ordered field lines
at the Alfven velocity for a time $dt$ years. The corresponding angular
scale is $\sim 2 \times 10^{11} \times dt$ km  ($0.4~dt''$).  Assuming
that the size of the smallest features resolved in Cas A at 15 GHz
(Arendt \& Dickel 1987) corresponds to the shortest scale length of the
ordered magnetic field, then $dt \sim$ 1 year. In the Cas A lifetime,
$\sim$ 300 years, electrons can diffuse over $\sim \sqrt{300/dt}
\times 0.4'' \times dt$ $\sim$ $7''$ $\times$ $\sqrt{dt}$.  Fluid
turbulence will also spread out locally accelerated electrons further.
Thus, we might expect to see spectral index variations on angular
scales of $\sim$ $7''$ and larger if the particles were accelerated
under differing physical conditions.  Magnetic field enhancements from
shocks and turbulence will create bright emission features which we see
as knots and filaments (AR96).  These features will have a spectral
index corresponding to the underlying relativistic electron
population.  The magnetic field is not sufficiently coherent to allow
diffusion of locally accelerated electrons over the entire remnant.
Given major azimuthal variations in the dynamics (Koralesky et al.,
1998), we are left with the conundrum of how the spectral index and
apparently the electron distribution can be so similar over the large
remnant.  The same thing occurs in much larger objects, such as IC443.




\section{Summary}

1) The images of Cas A show the same structures at 1.5, 5, 28, and  83 GHz at $7''$ resolution.

2) The average spectral index for the 26 brightest peaks is -0.82 +/-
0.05, somewhat steeper on average than the integrated flux density
(-0.77).

3) The spectra derived from images between 1.5 and 83 GHz show a range
of spectral indices from -0.75 to -0.95.  The data are most consistent
with power laws of different slopes rather than curved spectra.
The spectra between 1.5 and 83 GHz are consistent with no spectral
break below 83 GHz in the clumps.



4)  The five compact features outside the shell show significantly
steeper spectra.  The existence of different power laws at different
locations demonstrates that particle acceleration varies across the
remnant.  Diffusion can smooth the electron distributions over scales
on the order of our resolution (0.1 pc), but not throughout the entire
remnant.

\acknowledgements

This work was supported in part by NSF Grant AST-9613998 to the
University of California, and AST96-13999  to the University of
Illinois.  SNR research at the University of Minnesota is supported by
the NASA Graduate Research Program and the National Science Foundation
under grant AST 96-19438. We thank Mark Holdaway, Tom Jones, and an
anonymous referee for comments which have improved the presentation of
this work.

\clearpage
\newpage

\begin{table}
\centering
\caption{Observations}
\smallskip
\begin{tabular}{ccccccccc}
\hline

 DATE	& Frequency (GHz)	& Telescope	& uv-coverage (klambda)   \\ \hline
\hline
96SEP18	& 87		&	NRAO 12m		& 0 - 3.1	\\
96OCT16	& 83.1 - 86.6	&	BIMA 9-antenna c-array  & 1.9 - 36.3	\\
97JAN18	& 83.1 - 86.6	&	BIMA 9-antenna b-array  & 5.8 - 72.6	\\
97SEP02	& 28.5		&	BIMA 9-antenna c-array  & 0.6 - 6.7	\\
97SEP06 & 28.5  	&       BIMA 9-antenna b-array  & 1.8 - 26.5	\\
97NOV29 & 75.2 - 78.7	&       BIMA 10-antenna c-array & 1.5 - 20.7	\\
\hline
\end{tabular}
\raggedright\smallskip
\end{table}

\clearpage
\newpage
 
\begin{table}
\centering
\caption{Fitted Spectra}
\smallskip
\begin{tabular}{ccccccccc}
\hline
Feature & Flux density at 1 GHz\tablenotemark{a} & Spectral Index\tablenotemark{b} \\ \hline
\hline
A & 1.84 & -0.83 \\
B & 2.77 & -0.83 \\
C & 4.03 & -0.95 \\
D & 2.07 & -0.85 \\
E & 5.07 & -0.89 \\
F & 3.77 & -0.75 \\
G & 2.32 & -0.83 \\
H & 1.73 & -0.76 \\
I & 3.31 & -0.78 \\
J & 3.76 & -0.78 \\
K & 3.23 & -0.76 \\
L & 2.13 & -0.82 \\
M & 3.90 & -0.79 \\
N & 1.97 & -0.79 \\
O & 2.34 & -0.89 \\
P & 2.11 & -0.81 \\
Q & 6.07 & -0.85 \\
R & 4.95 & -0.92 \\
S & 4.07 & -0.86 \\
T & 2.67 & -0.79 \\
U & 2.26 & -0.85 \\
V & 1.71 & -0.75 \\
W & 2.83 & -0.78 \\
X & 3.44 & -0.78 \\
Y & 2.75 & -0.81 \\
Z & 1.32 & -0.78 \\
\hline
\end{tabular}
\tablenotetext{a}{The errors in the flux density at 1 GHz are around 10 mJy.}
\tablenotetext{b}{The errors in the spectral index are estimated both from alternate measurements of the
flux densities (see text), and from different weighting of the 4 frequencies range from 0.02 to 0.05.}
\raggedright\smallskip
\end{table}

\newpage
\clearpage

\section{Figure Captions}

Figure 1. The maximum entropy image at 83 GHz convolved with a beam FWHM
= $6.5 \times 6.2 ''$. This image contains the single dish data, and
samples spatial frequencies from 0 to 72590 wavelengths.  In this image
and the subsequent maps, the coordinates are relative to the center of
Cas A at RA 23:23:25.000 and Dec 58:49:00.000 (J2000).

Figure 2. The maximum entropy images at 1.5, 5, 28, and 83 GHz all
convolved to a common resolution $7''$, and scaled to the mean spectral
index over the image.  The contour interval is 33 mJy/beam at 83 GHz.
The synthesised beam FWHM is indicated by the filled ellipse in the
lower left of each panel.  The same structures are apparent in each
images.

Figure 3. The difference between pairs of images after scaling to the
mean spectral index over the image.  (Top left 83-5 GHz, top right
83-28 GHz, bottom left 28-5 GHz, bottom right 83-1.5 GHz).  The contour
interval is 12 mJy/beam at 83 GHz.  The synthesised beam FWHM is
indicated by the filled ellipse in the lower left of each panel.
Significant large scale differences are apparent. These may be a result
of instrumental effects such as different uv-sampling, primary beam and
pointing errors, and image processing errors.

Figure 4. Maximum entropy images at 1.5, 5, 28, and 83 GHz each
spatially filtered to the same common uv-range defined by a circular
annulus from 1870 to 26500 wavelengths. The effective convolving beam
includes structures from $7''$ to $95''$.  The images are scaled to the
mean spectral index over the image.  The contour interval is 21
mJy/beam at 83 GHz.  The synthesised beam FWHM is indicated by the
filled ellipse in the lower left of each panel.

Figure 5. The difference between pairs of spatially filtered images,
after scaling to the mean spectral index over the image.  (Top left
83-5 GHz, top right 83-28 GHz, bottom left 28-5 GHz, bottom right
83-1.5 GHz).  The contour interval is 9 mJy/beam at 83 GHz.  The
synthesised beam FWHM is indicated by the filled ellipse in the lower
left of each panel.  There are some common features on these images
which we believe correspond to real spectral index variations across
the source.

Figure 6. Profiles in position angles 0 and 90 degrees of the measured
voltage primary beam pattern using holography on a transmitter at 90.4
GHz (solid line). The dashed line shows the Gaussian model used in the
image reconstruction, and the dot-dashed line shows the differences
between the measurement and the model.

Figure 7. The locations of 26 features which are readily identifiable
at 1.5, 5, 28 and 83 GHz at $7''$ resolution. The contours show the
spatially filtered image at 5 GHz. The contour interval corresponds to
240 mJy/beam at 83 GHz when scaled by a spectral index -0.77.

Figure 8. Spectra of the 26 features shown in figure 7 (two panels).
The flux densities are scaled by a spectral index -0.77, and normalized
to 83 GHz.  The vertical, linear flux density scale is indicated at the
bottom.  The errors at each frequency are indicated at the top of the
panel. Although most of the spectra are straight within the errors, the
steepest spectra show a somehat concave shape (see section 5.2).

Figure 9. Color-color plots 1.5, 5 and 83 GHz.  The solid line is the
locus of power laws;  2 power law indicies are labeled for
illustration.  The dotted line shows a continuous injection model with
a low-frequency index of -0.75. (see text)

\newpage

\clearpage

\begin{figure}
\plotone{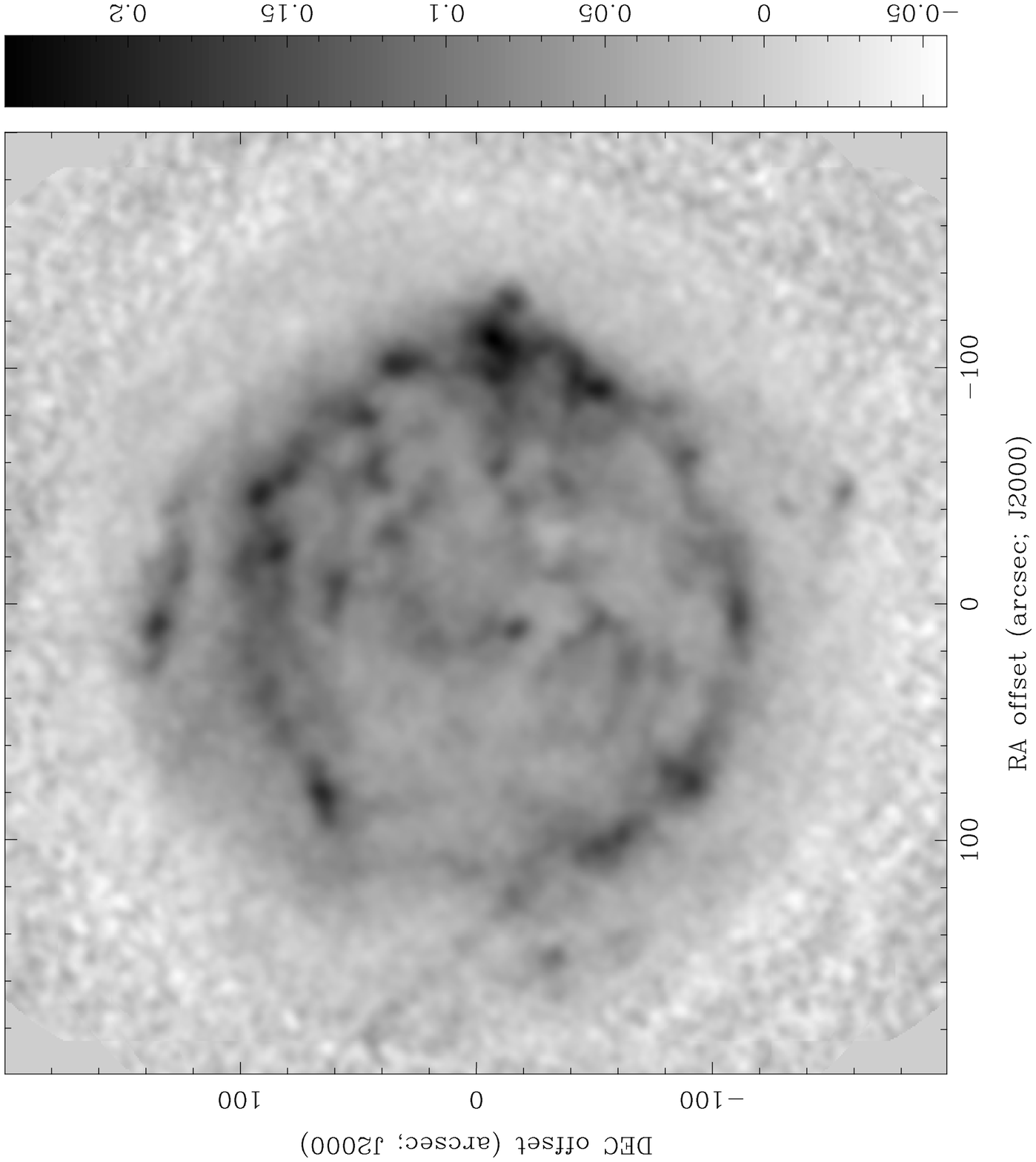}
\end{figure}

\begin{figure}
\plotone{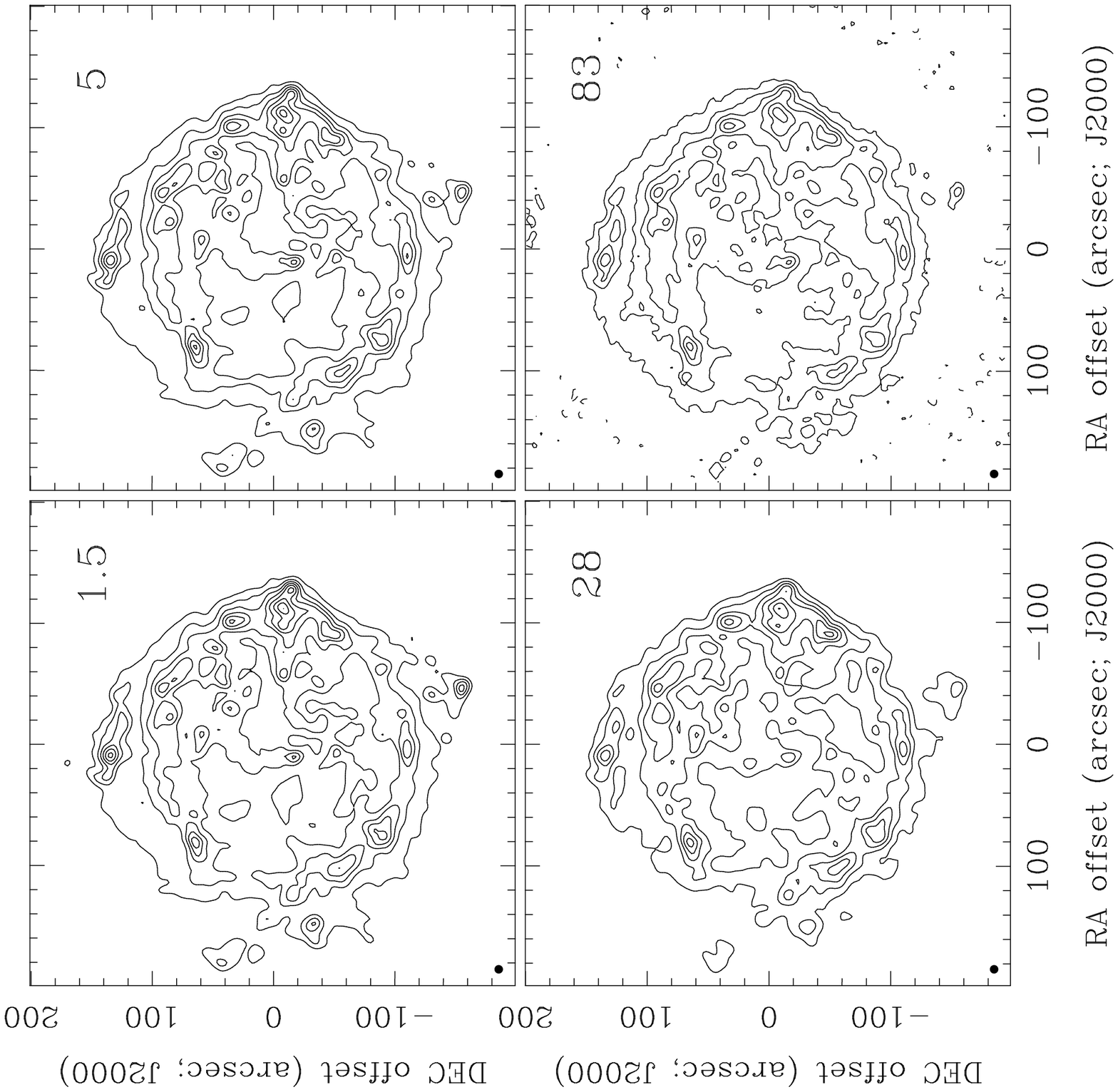}
\end{figure}

\begin{figure}
\plotone{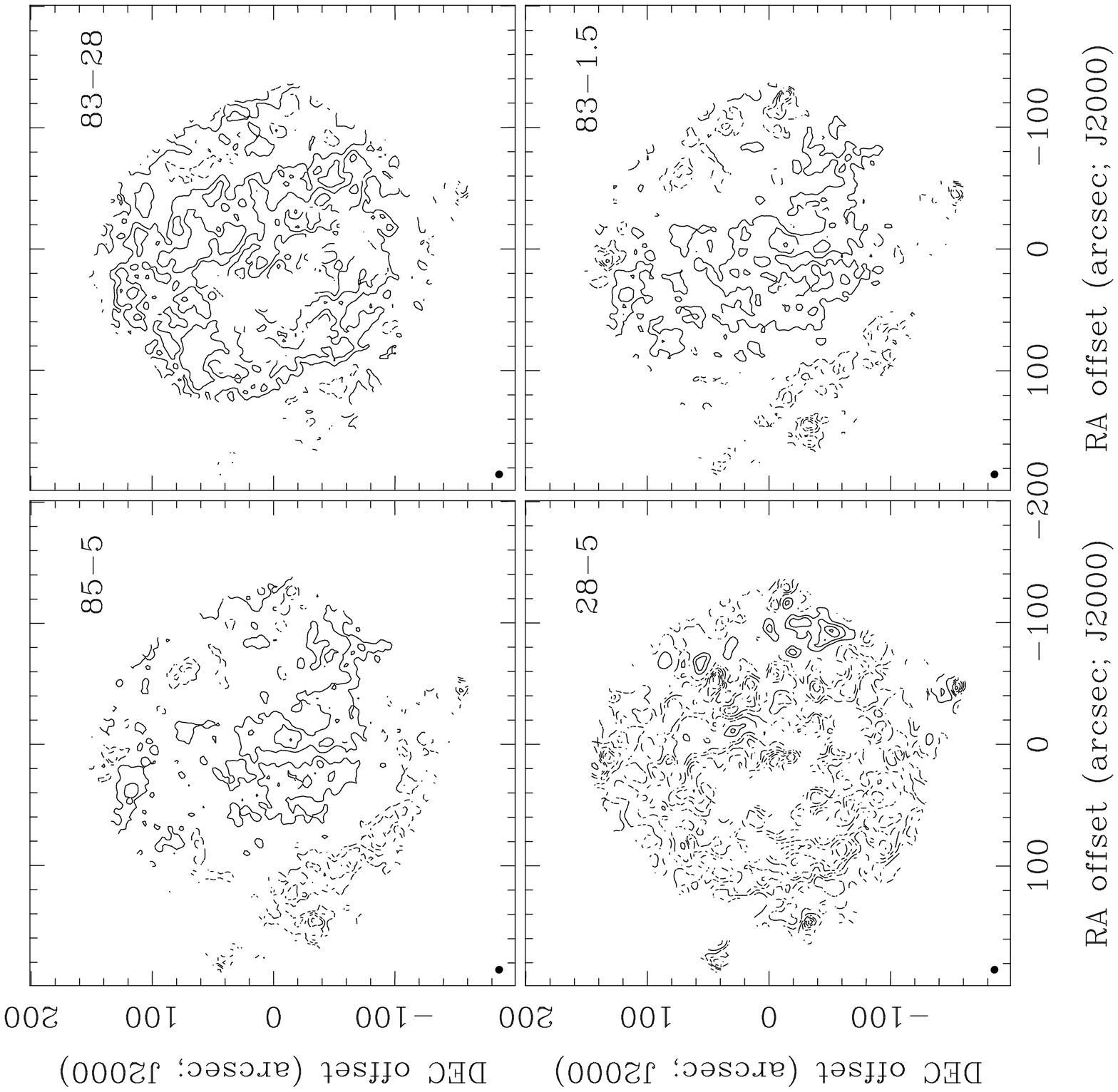}
\end{figure}

\begin{figure}
\plotone{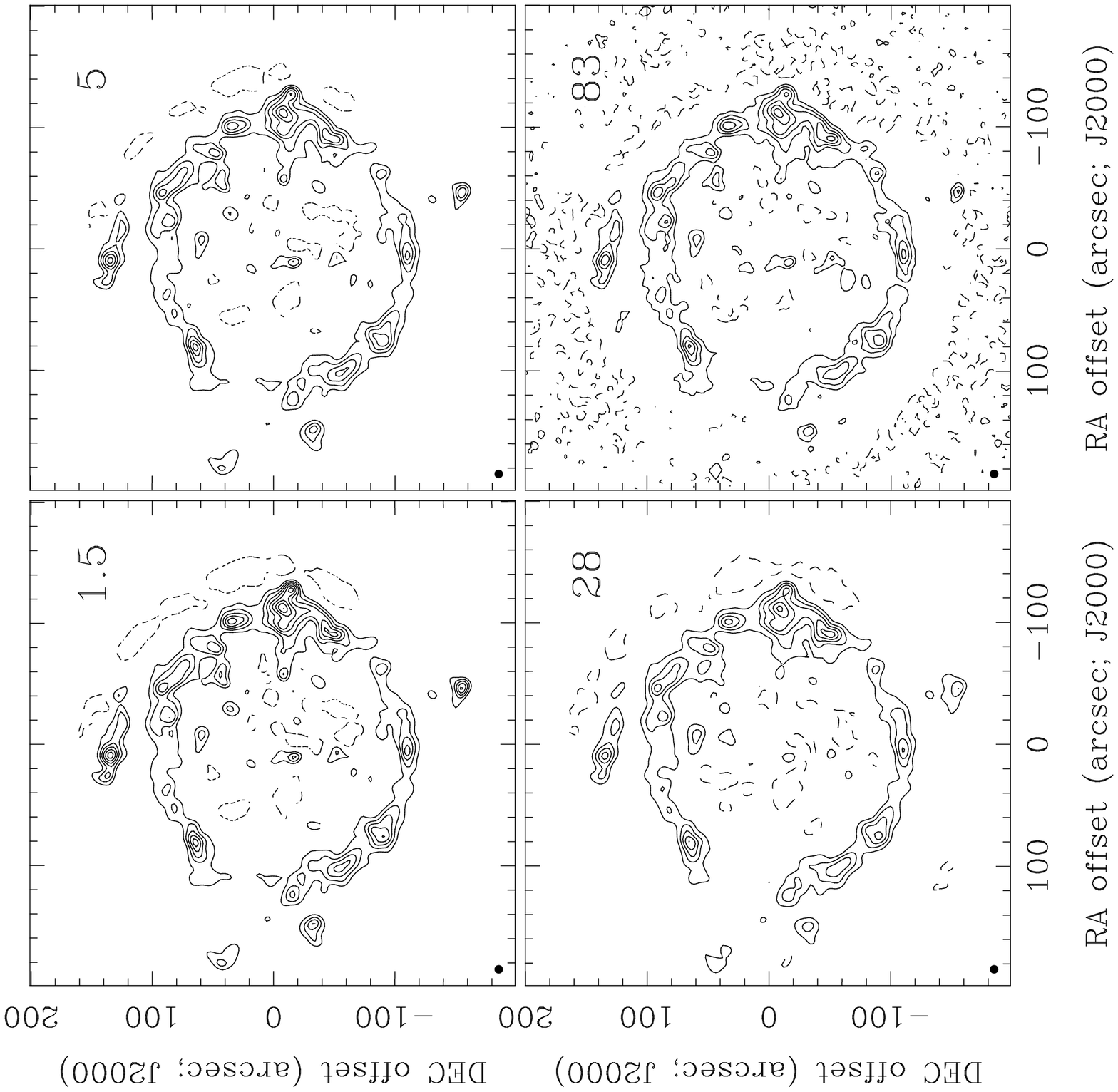}
\end{figure}

\begin{figure}
\plotone{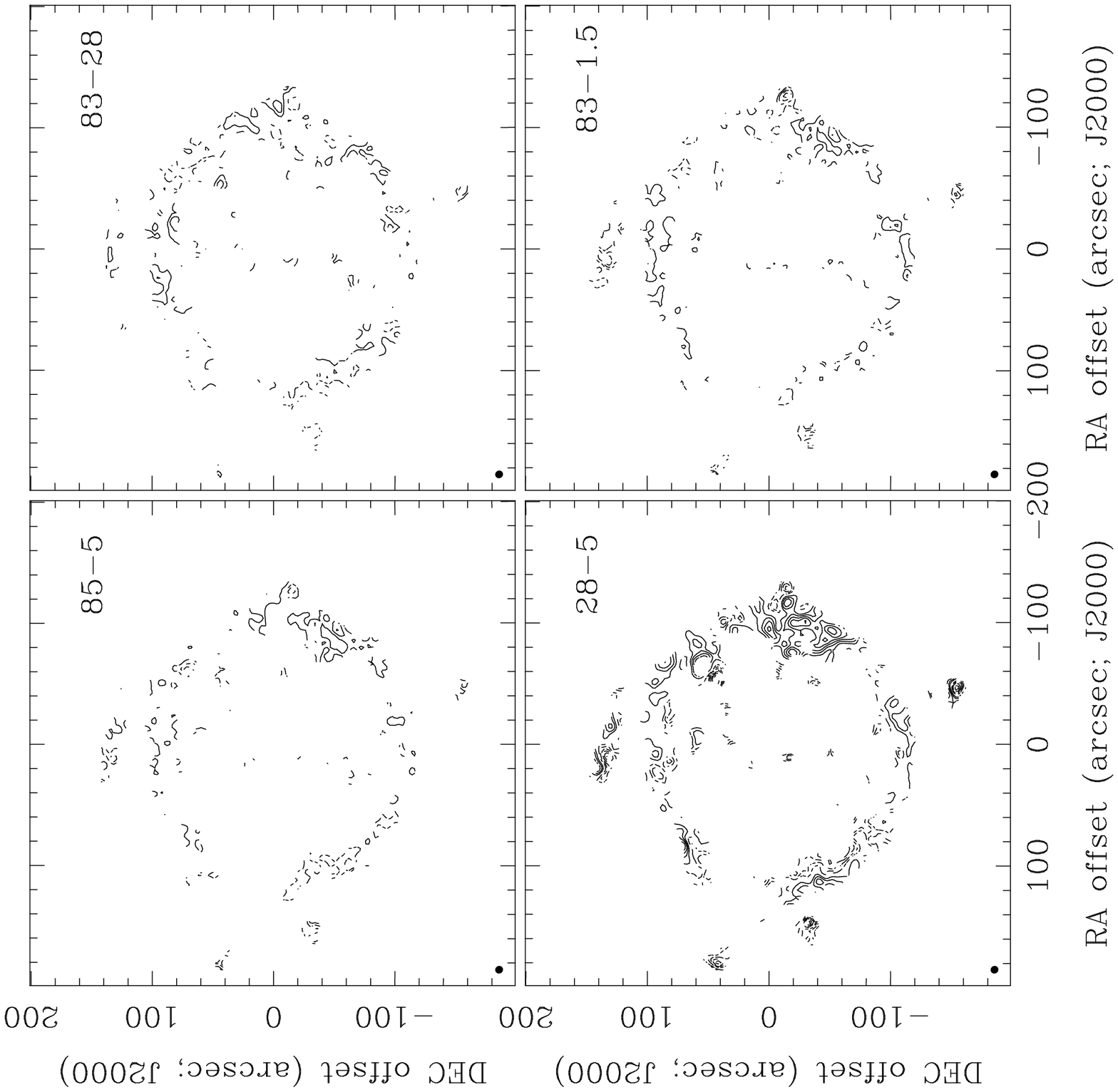}
\end{figure}

\begin{figure}
\plotone{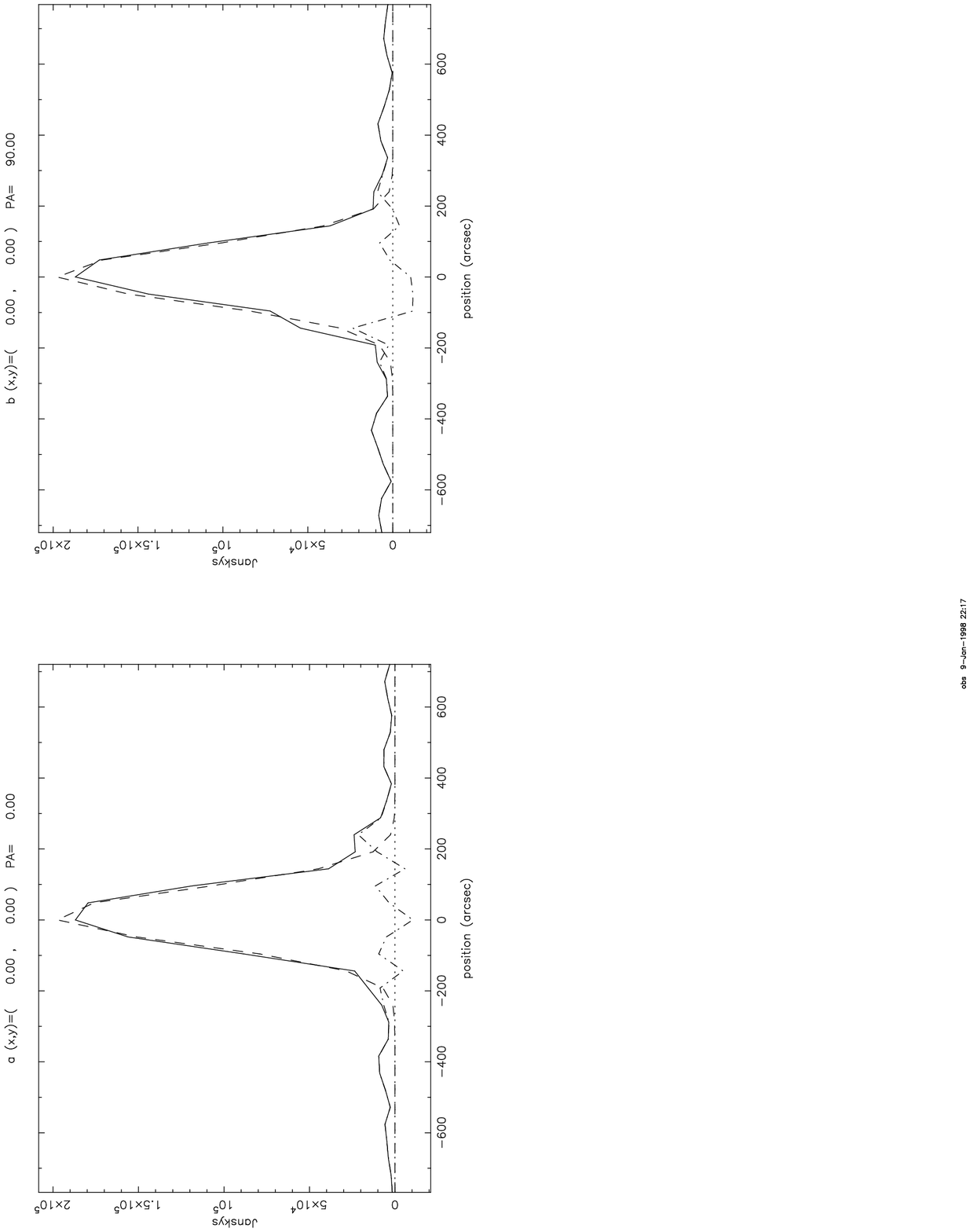}
\end{figure}

\begin{figure}
\plotone{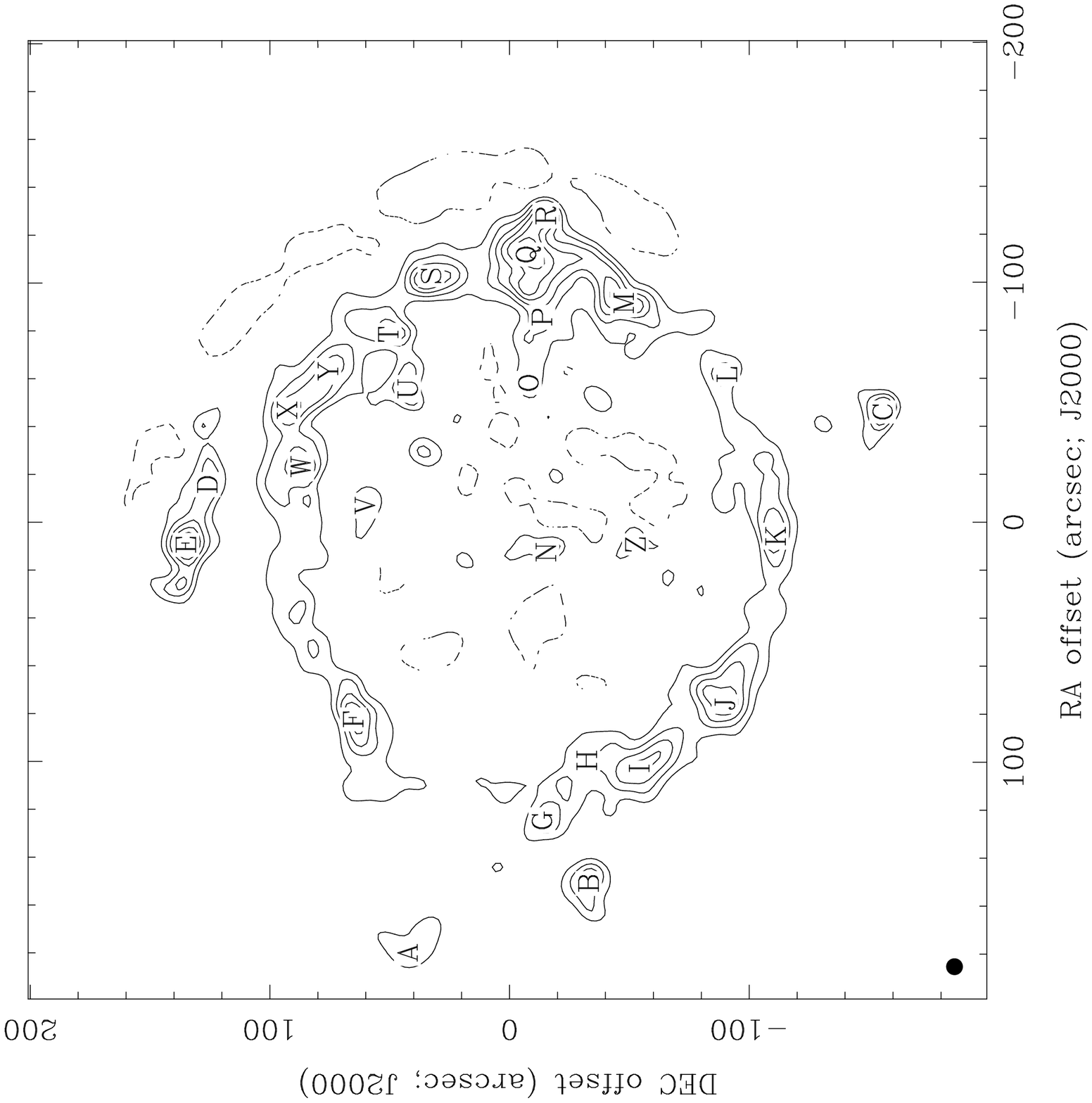}
\end{figure}

\epsscale{.5}

\begin{figure}
\plotone{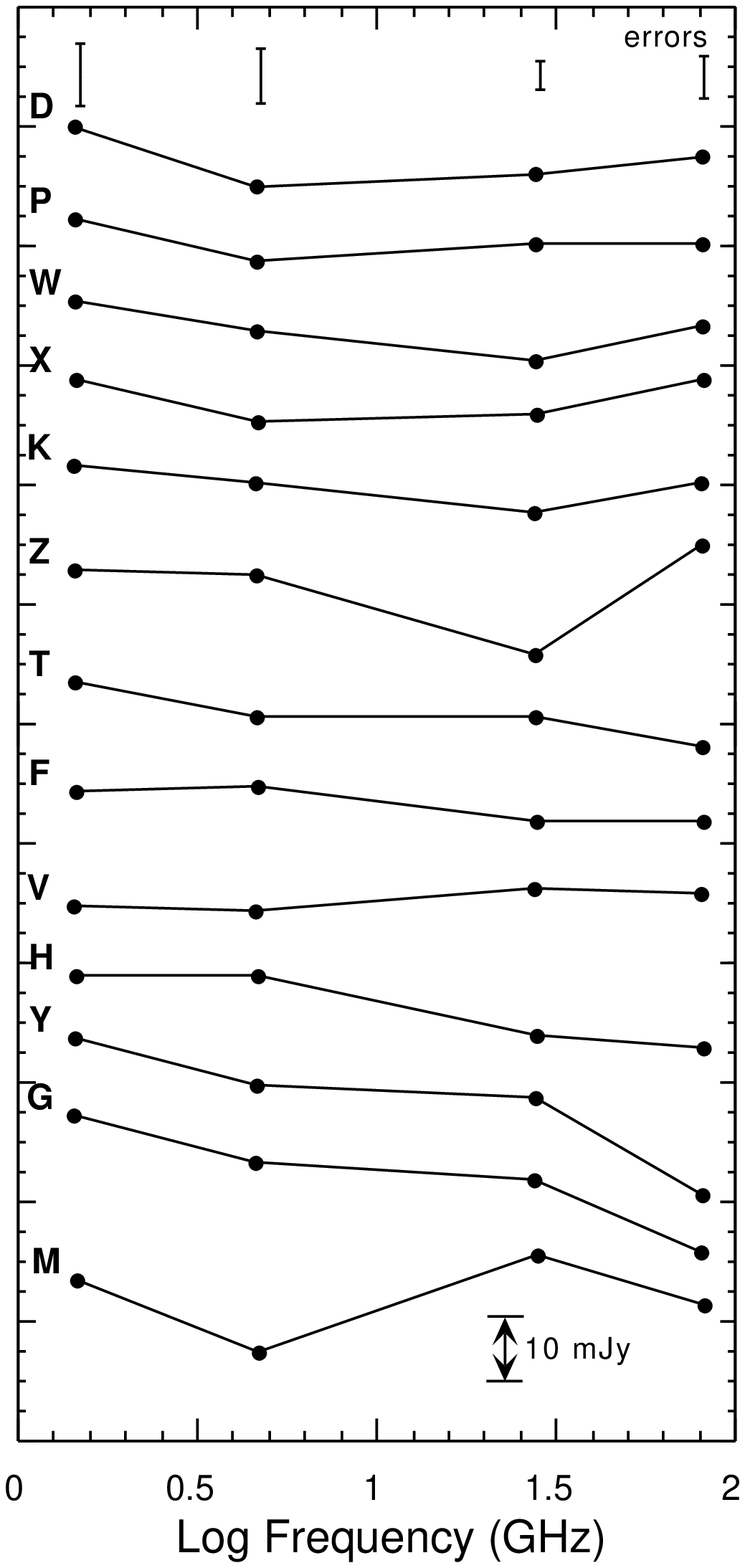}
\end{figure}

\begin{figure}
\plotone{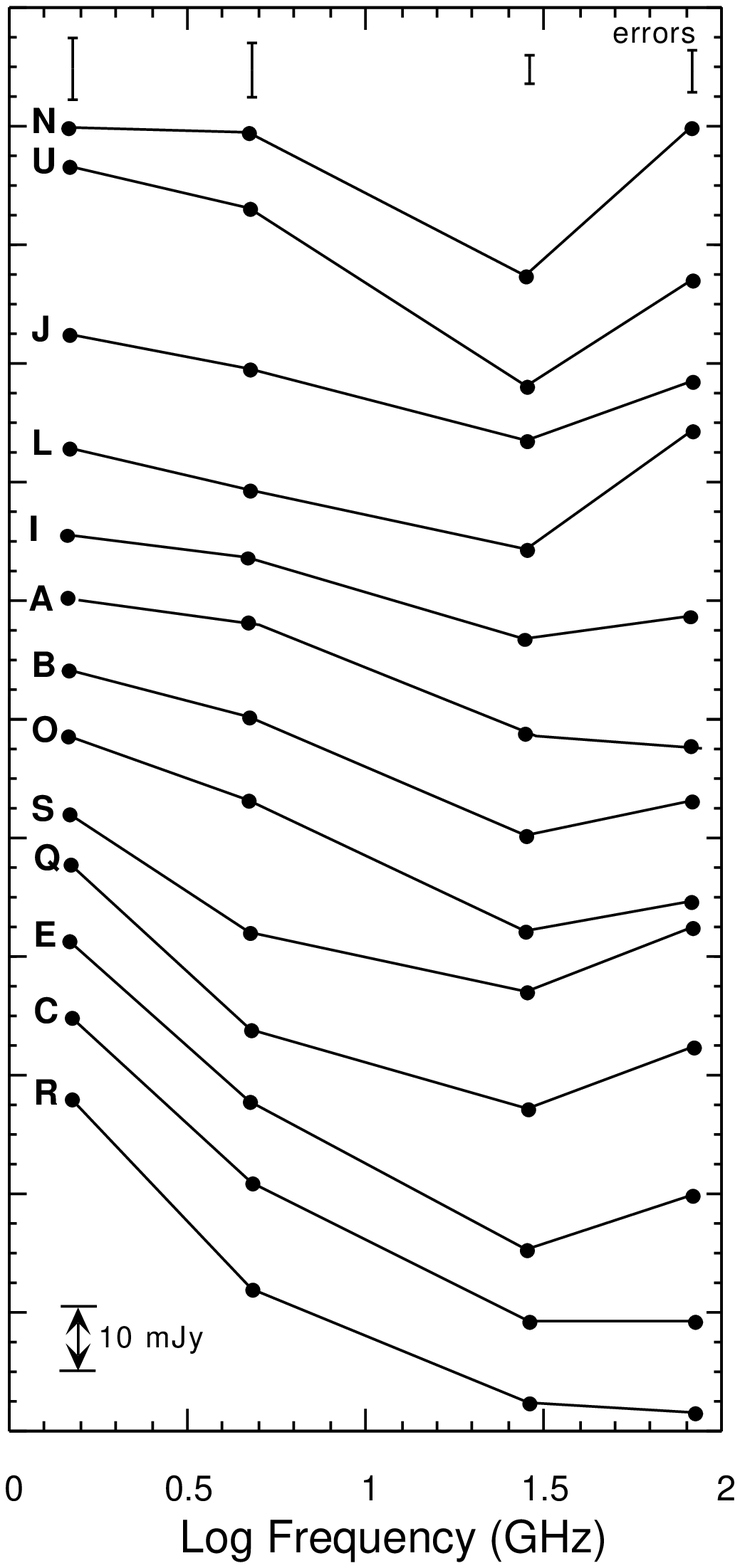}
\end{figure}

\epsscale{1}

\begin{figure}
\plotone{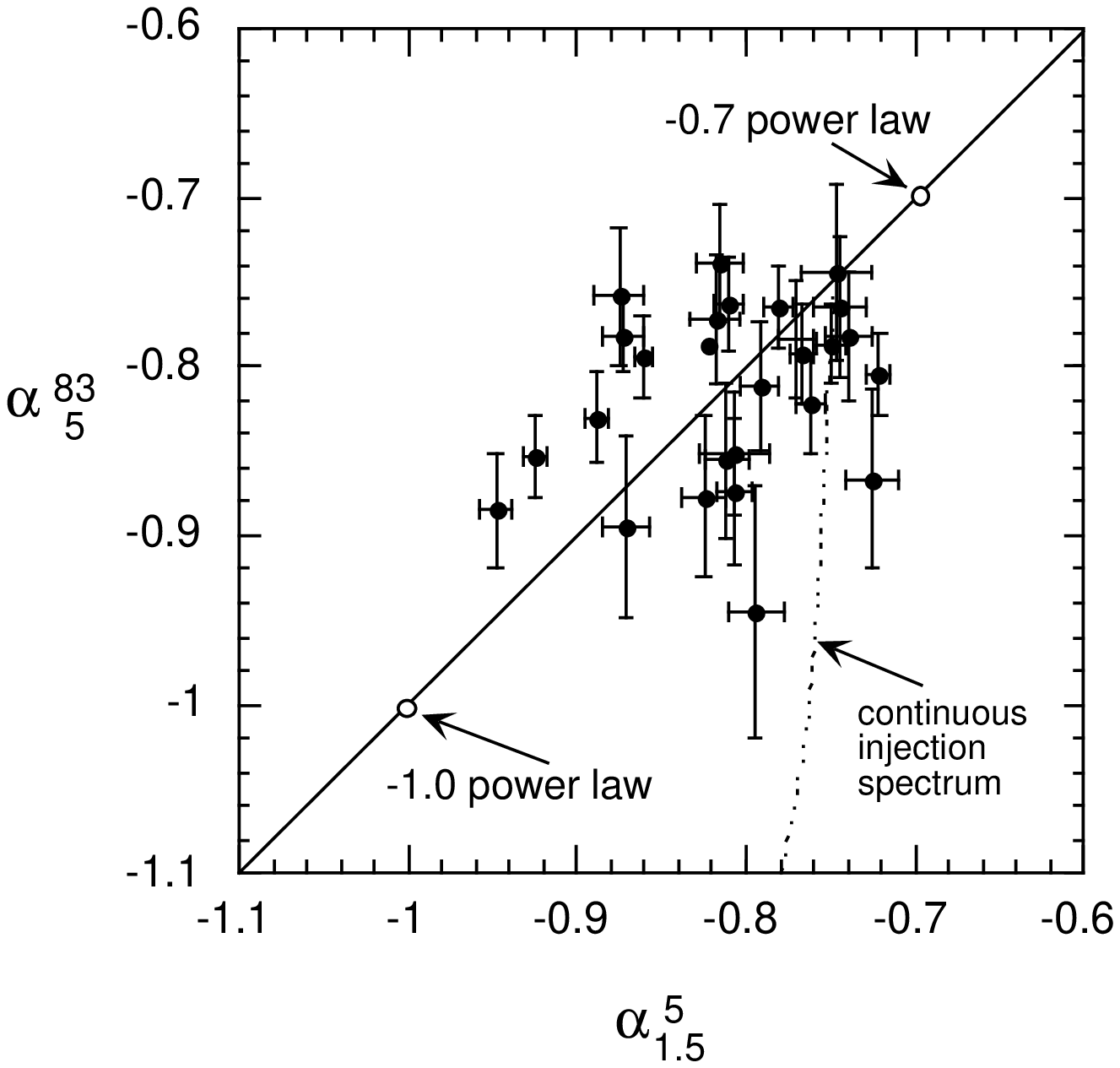}
\end{figure}

\end{document}